\documentclass{ws-mplb}
\usepackage{amsfonts,amssymb,amsmath}
\usepackage{graphicx}
\usepackage[super,sort,compress]{cite} 

\begin{document}

\markboth{R. Fr\'esard and V. H. Dao}{Charge instabilities of the extended attractive Hubbard Model on the cubic lattice}

%
\catchline{}{}{}{}{}
%

\title{Charge instabilities of the extended attractive Hubbard Model on the cubic lattice}

\author{\footnotesize Raymond Fr\'esard}

\address{Normandie Univ, ENSICAEN, UNICAEN, CNRS, CRISMAT, 14050 Caen, France\\
Raymond.Fresard@ensicaen.fr}

\author{Vu Hung Dao}

\address{Normandie Univ, ENSICAEN, UNICAEN, CNRS, CRISMAT, 14050 Caen, France\\
vu-hung.dao@ensicaen.fr}

\maketitle

\begin{history}
\received{(Day Month Year)}
\revised{(Day Month Year)}
\end{history}

\begin{abstract}
The paramagnetic phase of the extended attractive Hubbard model on the cubic
lattice is studied within the spin rotation invariant Kotliar-Ruckenstein
slave-boson representation at zero temperature. It is obtained that the
quasiparticle residue of 
the Fermi liquid phase vanishes for all densities at an interaction strength
slightly smaller than $U_c$ that signals the Brinkman-Rice transition, and
that it weakly depends on density. While for vanishing non-local interaction
parameters homogeneous static charge instabilities are found in a rather
narrow window centered around quarter filling and $U \simeq 0.8\ U_c$,
increasing them to $V = -0.2\ U$ results into a severe narrowing of this
window. On the contrary, when all interaction parameters are attractive, for
example for $V = 0.2\ U$, a large parameter range in which homogeneous static
charge instabilities is found. Yet, this systematically happens inside the
Fermi liquid phase. 
\end{abstract}


\section{Introduction}
While the Hubbard model has been originally introduced to describe metallic
magnetism \cite{Hub63,Gut63,Kan63}, it gained renewed interest after Anderson's
proposal that it represents a minimal model for the $d$ electrons within the
CuO$_2$ layers common to the high T$_c$ superconductors \cite{And87}. Yet, it
is fair to say that this model still lacks a broadly accepted solution,
especially in the strongly correlated regime, despite its apparent simplicity
following from its assumption that the non-local Coulomb interaction is fully
screened but locally. More recently, it has been proposed that non-local
interactions are of relevance, too, especially in the context of
two-dimensional systems such as surface systems\cite{Han13}, graphene
\cite{Sch13}, transition metal dichalcogenides\cite{vanL17}, ultracold
fermions in optical lattices\cite{Dut13,Dut15,vanL15}, and correlation induced
capacitance enhancement\cite{Stef17}. 

While the local interaction is repulsive in most instances, attractive
interaction has been invoked in ultracold fermion experiments \cite{Ess10} and
also in the condensed matter context \cite{vdM88,Mic90} and received strong
interest \cite{vanLo18,Ama10,Air13,Ter17}. In particular, in the locally
repulsive but non-locally attractive case, phase separation into a high and
low density state has been reported \cite{Fres16,Stef17}. 

The purpose of this work is to establish the influence of an attractive local
interaction on this instability in general, and on the Landau Fermi liquid
parameter $F_0^s$ in particular. We hence focus on the extended nondegenerate
Hubbard model which describes the simplest correlated metals and investigate
the influence of intersite Coulomb on the charge instabilities. Thereby we
follow the route initiated by Vollhardt \cite{Vol84} who investigated Landau
Fermi liquid parameters $F_0^a$ and $F_0^s$ for $^3$He. Further studies gave
the Fermi liquid interaction and the quasiparticle scattering amplitude on the
Fermi surface in $^3$He \cite{Pfi86}. 

Since we are also interested in the strongly correlated regime we perform our
investigations in a framework which is able to 
capture interaction effects beyond the physics of Slater determinants.
It is an extension of the Kotliar and Ruckenstein slave boson
representation that reproduced the Gutzwiller approximation on the
saddle-point level \cite{Kot86} and entails the interaction driven
Brinkman-Rice metal-to-insulator transition \cite{Bri70}. A whole range
of valuable results have been obtained with Kotliar and Ruckenstein
\cite{Kot86} and related \cite{Li89,FW} slave boson representations
which motivate the present study.
In particular they have been used to describe
antiferromagnetic \cite{Lil90}, ferromagnetic \cite{Doll2}, spiral
\cite{Fre91,Arr91,Fre92} and striped \cite{SeiSi,Sei02,Rac06a,Rac07a}
phases. Furthermore, the competition between
the latter two has been addressed as well \cite{RaEPL}. Besides, it has been
obtained that the spiral order continuously evolves to the ferromagnetic order
in the large $U$ regime ($U \gtrsim 60t$)~\cite{Doll2} so that it is unlikely
to be realized experimentally. Consistently, in the two-band model,
ferromagnetism was found as a possible groundstate only in the doped Mott
insulating regime~\cite{Lamb02}. Yet, adding a ferromagnetic exchange coupling
was shown to bring the ferromagnetic instability line into the intermediate
coupling regime~\cite{Lhou15}. A similar effect has been obtained
with a sufficiently large next-nearest-neighbor hopping amplitude~\cite{FW98} or
going to the fcc lattice~\cite{Igo15}. The influence
of the lattice geometry on the metal-to-insulator transition was
discussed, too \cite{Kot00}. For instance, a very good agreement with
Quantum Monte Carlo simulations on the location of the
metal-to-insulator transition for the honeycomb lattice has been
demonstrated \cite{Doll3}. Finally, further motivation comes from
the strongly inhomogeneous polaronic states that have been found in correlated
heterostructures using the Hubbard model extended with intersite
Coulomb interactions \cite{Pav06}.  

Furthermore, comparison of ground state energies to existing numerical
solutions have been carried out for the square lattice, too. For
instance, for $U=4t$ it could be shown that the slave boson ground
state energy is larger than its counterpart by less than 3\%
\cite{Fre91}. For larger values of $U$, it has been obtained that the
slave boson ground state energy exceeds the exact diagonalization data
by less than 4\% (7\%) for $U=8t$ ($20t$) and doping larger than 15\%.
The discrepancy increases when the doping is lowered \cite{Fre92}. It
should also be emphasized that quantitative agreement to quantum Monte
Carlo charge structure factors was established \cite{Zim97}.

The purpose of this paper is to evaluate the Fermi liquid
Landau parameter $F_0^s$ for the metallic state in the extended Hubbard model
in the case where the local interaction is attractive. This extends earlier
work on the locally repulsive case\cite{Fres16} that lead to electronic phase
separation which may considerably enhance the capacitance of capacitors when
their metallic plates are strongly correlated\cite{Stef17}. 

\section{Extended Hubbard model}
\label{sec:hub}

Numerous studies of correlated electrons have been devoted to the
properties of the Hubbard model on a square lattice, especially after
Anderson's proposal that it represents a minimal model for the $d$
electrons within the CuO$_2$ layers common to the high T$_c$
superconductors \cite{And87}. Yet the Hubbard model assumes a perfect
screening of the long-range part of the Coulomb interaction. This may
be questionable and the relevance of this approximation may be
assessed by considering the extended Hubbard model that reads:
\begin{equation}
H = \sum_{i,j,\sigma}t_{ij}c_{i\sigma}^{\dagger}c_{j\sigma}
+ U\sum_{i}n_{i\uparrow} n_{i\downarrow} +
\frac12 \sum_{i,j}V_{ij}n_{i} n_{j}
\label{eq:model}
\end{equation}
and includes intersite Coulomb $V_{ij}$ 
interactions. These elements decay fast with increasing distance
$|\vec{R}_i-\vec{R}_j|$, but extend in general beyond nearest
neighbors. Here $c_{i\sigma}^\dagger$ are electron creation
operators at site $i$ with spin $\sigma$,and
$n_{i\sigma}=c_{i\sigma}^{\dagger}c_{i\sigma}^{}$. We consistently use the
particle-hole symmetric form for both density-density interaction terms.
Although one expects that $V_{ij}>0$, in certain cases effective
intersite Coulomb interactions may be attractive \cite{Mic90}.
Therefore, we shall treat $\{V_{ij}\}$ as effective parameters and
consider both signs of them below. 

We perform our calculations in the spin rotation invariant formulation of the
Kotliar-Ruckenstein slave boson representation of the Hubbard model
\cite{Kot86,FW,Lhou15}. Though such functional integrals can be calculated
exactly for the Ising chain \cite{Fre01} and some toy models \cite{Kop07},
even with the Kotliar and Ruckenstein representation \cite{Kop12,Dao19}, this
is unpractical on higher dimensional lattice. Here we rather resort to a
calculation at one-loop order around the paramagnetic saddle-point. This
saddle-point approximation is exact in the large degeneracy limit, and the
Gaussian fluctuations provide the $1/N$ corrections \cite{FW}. Moreover it
obeys a variational principle in the limit of large spatial dimensions where
the Gutzwiller approximation becomes exact for the Gutzwiller wave function
\cite{Met89a,Met89b,Met88}. Furthermore, it could be shown in this limit that
longer ranged interactions are not dynamical and reduce to their Hartree
approximation \cite{Mul89}. Therefore, our approach also obeys a variational
principle in this limit when applied to the above extended Hubbard model
Eq.~(\ref{eq:model}). 
All used formulas may be found in Lhoutellier et. al.'s work \cite{Lhou15}.

\section{Saddle-point solutions}

\begin{figure}
   \centering
\begin{tabular}{cccc}
\includegraphics[clip=true,trim={3cm 0 2.4cm 0}, width=0.251\textwidth]{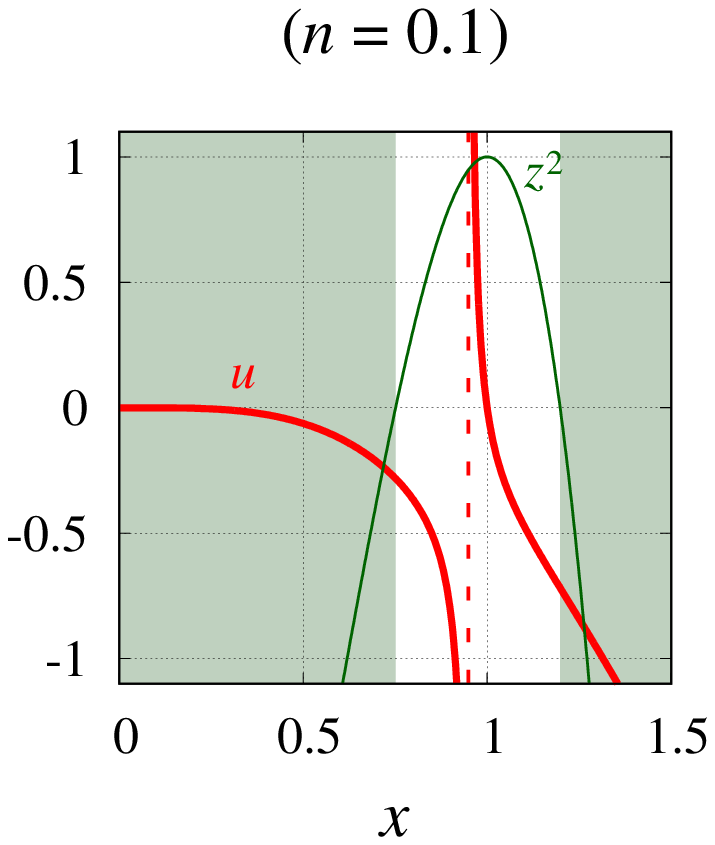}&
\includegraphics[clip=true,trim={4cm 0 2.6cm 0}, width=0.21\textwidth]{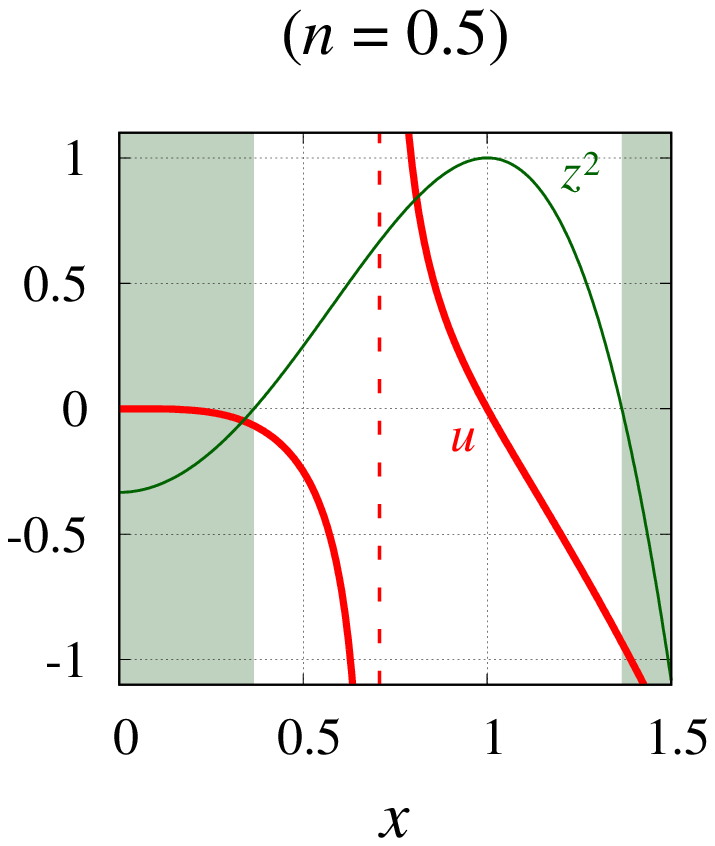}&
\includegraphics[clip=true,trim={4cm 0 2.6cm 0}, width=0.21\textwidth]{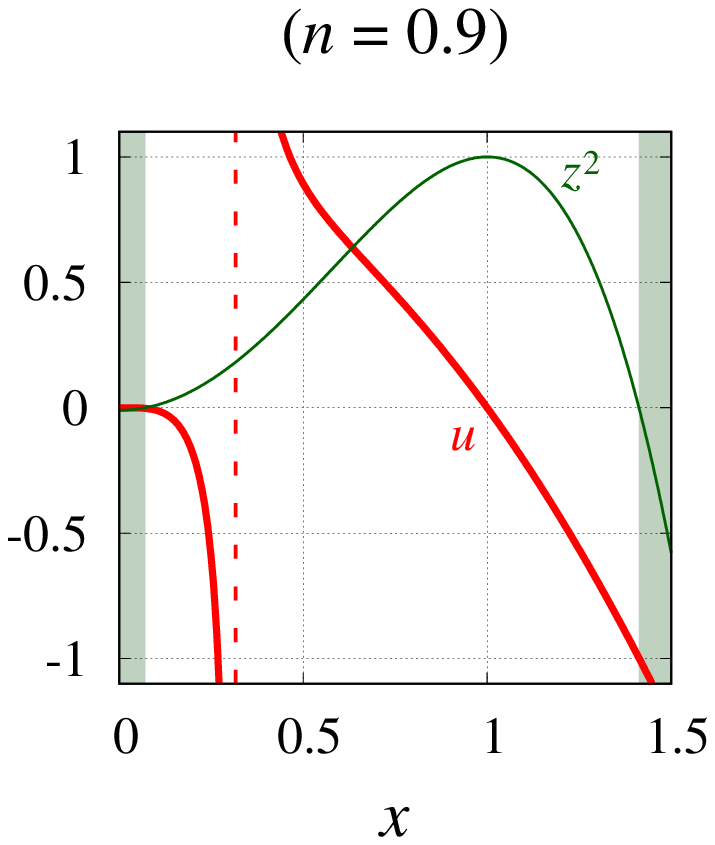}&
\includegraphics[clip=true,trim={4cm 0 2.6cm 0}, width=0.21\textwidth]{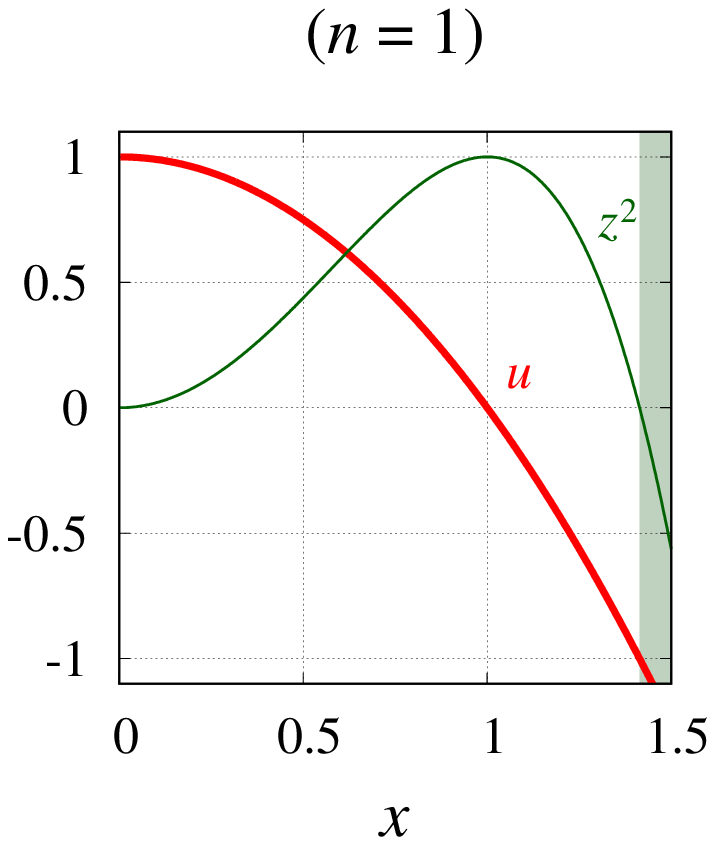}
\end{tabular}
    \caption{ Saddle-point function $u$ (red, thick line) and renormalization factor 
    $z^2$ (green, thin line) as functions of $x$, for different electron densities. 
    The shaded area corresponds to unphysical values of $x$ yielding $z^2 <0$.}
    \label{fig:u_x} 
\end{figure}

In the translational invariant paramagnetic phase, all local quantities
are site independent. The saddle-point values of the boson fields representing 
the atomic states with respectively zero, single and double occupancy may be 
expressed as  
\begin{equation}
 e = \frac{x^2 + \delta}{2x} ,
\quad p_0 = \sqrt{1 - \frac{x^4 + \delta^ 2}{2 x^2}}, 
\quad d = \frac{x^ 2 - \delta}{2 x},
\end{equation}
where the doping from half-filling is $\delta = 1 -n$, $n$ is the electron 
density, and the variable $x =e +d$. 
Hence the auxiliary boson fields obey the physical constraints
on average~:
\begin{equation}
e^2 + p_0^2  + d^2 = 1, 
\quad {\rm and} \quad p_0^2 + 2 d^2 = 1- \delta.
\end{equation} 

The value of $x$ is determined by solving the saddle-point equation \cite{Vol87,FW,Dao17}
\begin{equation}\label{eq:saddle}
u(x) \equiv  \frac{(1-x^2) x^4}{x^4 - \delta^2}  =  \frac{U}{U_0}.
\end{equation}
Here we have introduced the coupling scale
\begin{equation}
 U_0 = - \frac{8}{1 - \delta^2} \tilde{\varepsilon}
\end{equation}
where $\tilde{\varepsilon}$ is the averaged kinetic energy. Here we do not
attempt to unravel the richness arising at finite temperature as known in the
repulsive case, \cite{Doll3,Dao17,Mez17} but we instead stick to zero
temperature.
Remarkably,
the intersite interaction $V_{i,j}$ does not enter Eq.~(\ref{eq:saddle}).
In the paramagnetic phase, it only influences the fluctuations, and
does not change the electron localization driven by the strong 
onsite interaction $U$.

The saddle-point equation can have zero, one, or two solutions, 
depending on the onsite Coulomb interaction and the density. 
The function $u(x)$ is plotted in Fig.~\ref{fig:u_x} for different values 
of the density, together with the inverse-mass renormalization factor
\begin{equation}
 z^2 = \frac{2 p_0^2 (e+d)^2}{1 -\delta^2} = 1 - \frac{(1-x^2)^2}{1-\delta^2}.
\end{equation} 

In the case of a positive $U$ there is at most one solution. For a finite doping 
the function $u(x)$ diverges at $x=\sqrt{|\delta|}$, so the saddle-point equation 
always possesses one solution, however large $U$ is. 
But at half-filling, $u(x)= 1-x^2$. As a consequence, the solution exists only 
below a critical value $U_c = \lim\limits_{\delta \rightarrow 0} U_0 = 
- 8\tilde{\varepsilon}$. For the presently investigated 3D cubic lattice 
$U_c = 16.0387t$. At this coupling the Brinkman-Rice transition occurs from a 
metallic state to a Mott insulator. Indeed the effective mass diverges as $z^2$ 
vanishes at $x=0$ for $U=U_c$, as shown in Fig.~\ref{fig:u_x}.

\begin{figure}
   \centering
\begin{tabular}{cc}
\includegraphics[clip=true,trim={1.2cm 0 0cm 0}, width=0.55\textwidth]{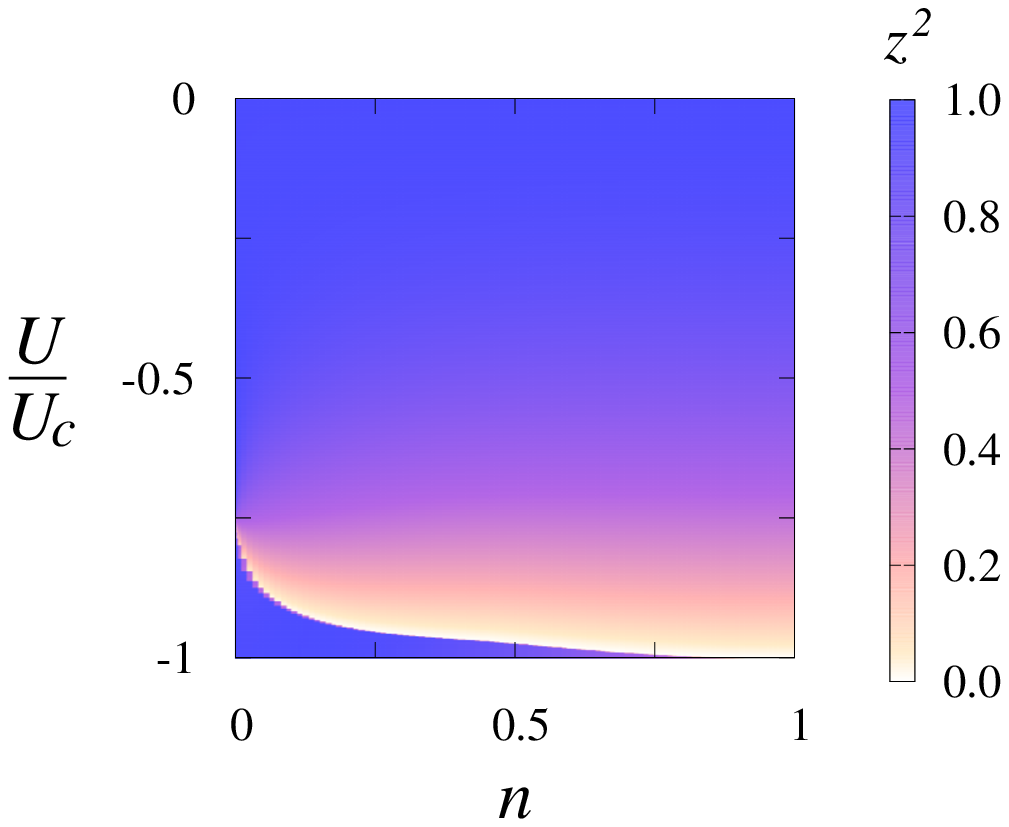}&
\includegraphics[clip=true,trim={3.3cm 0 0 0}, width=0.45\textwidth]{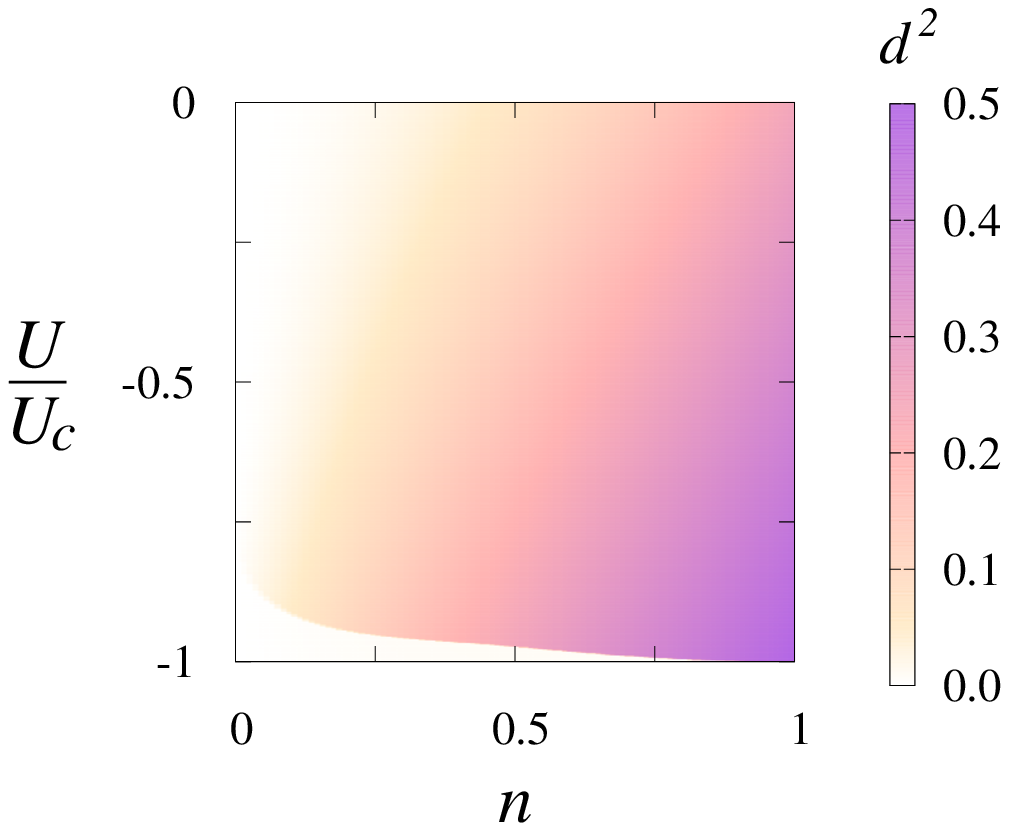}
\end{tabular}
    \caption{(left) The inverse-mass renormalization factor $z^2$ and (right) 
    the saddle-point value of the double occupancy $d^2$, as functions of the
    density $n$ and the onsite Coulomb interaction $U$.}
    \label{fig:d-z} 
\end{figure}

For a negative $U$ the saddle-point equation can have two solutions. But some
of them yield $z^2< 0$ and, thus, are not physical. The allowed values are
restricted to the range $\sqrt{1 -\sqrt{1-\delta^2}} \le x \le
\sqrt{1+\sqrt{1-\delta^2}}$. At half-filling, mirroring the positive $U$
situation, there is a single solution, which is relevant only for
$|U|<U_c$. At the critical coupling, the effective mass diverges and a
transition into an insulating state takes place. Beyond, the values
$x>\sqrt{2}$ are not physical. At density $n<1$, the saddle-point equation
possesses two solutions~: $x_-<\sqrt{|\delta|}$ and
$x_+>\sqrt{|\delta|}$. When $x_+$ is allowed, it is the saddle-point solution
with the lowest free energy~: it yields the largest double occupancy $d^2$
(hence the lowest interaction energy), and often, the largest $z^2$ (hence the
lowest kinetic energy) as well. Similarly to the half-filling case, the
solution $x_+$ is not physical for $|U|$ larger than a critical coupling
$\tilde{U_c}$. The latter decreases from $U_c$ at half-filling to $U_0/2$ at
$n=0$. As shown in Fig.~\ref{fig:d-z}, the renormalization factor $z^2$
decreases from unity at $U=0$ to zero at the critical coupling. Along the line
$\tilde{U_c}(n)$ the paramagnetic state described by the solution $x_+$ is an
incoherent insulator. For $|U|>\tilde{U}_c$, only the solution $x_-$ is
possible. However the latter yields a vanishingly small value of $d^2$ (see
Fig.~\ref{fig:d-z}). One can then doubt the relevance of this saddle-point
solution, because one would expect that the attractive onsite interaction
should favor the double occupancy. Since no plausible homogeneous phase may
be stabilized, one should seek for inhomogeneous ones, which however is
outside the scope of the present work. We will not discuss further 
the solution $x_-$ since, as shown below, the uniform paramagnetic state is
already unstable for values of $|U|$ smaller than the critical coupling.

\section{Charge instabilities}

\begin{figure}
   \centering
\begin{tabular}{cc}
\includegraphics[clip=true,trim={1.2cm 0 0cm 0}, width=0.55\textwidth]{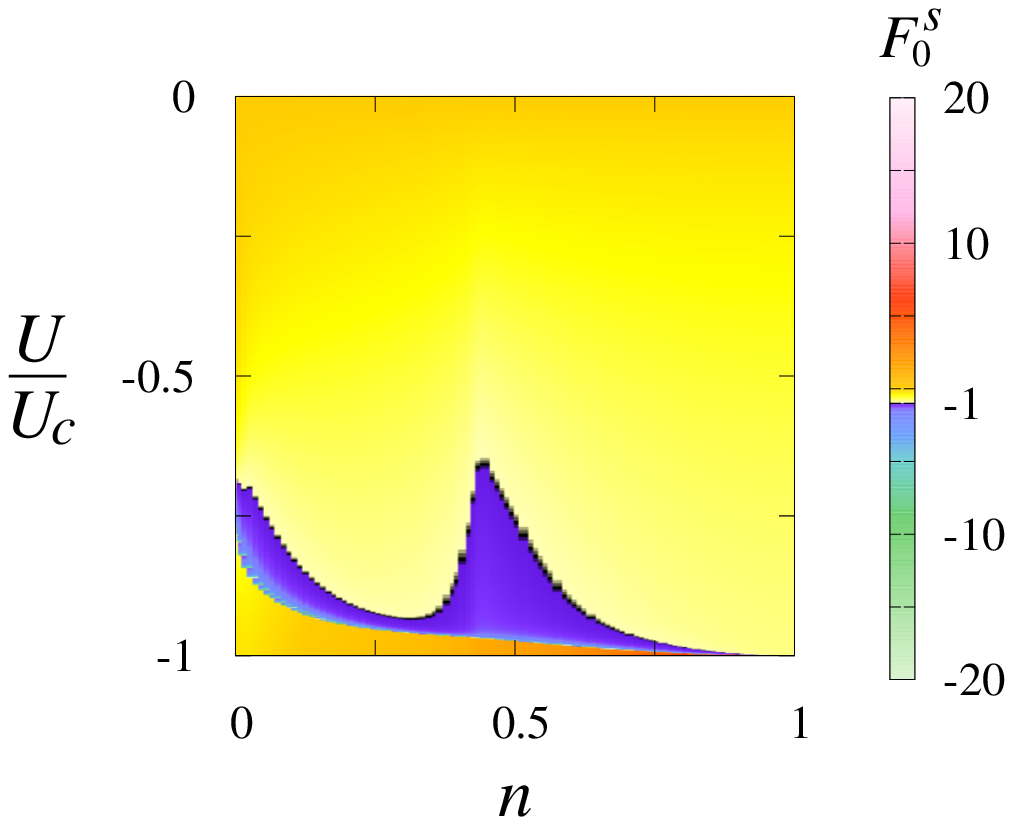}&
\includegraphics[clip=true,trim={3.3cm 0 0 0}, width=0.45\textwidth]{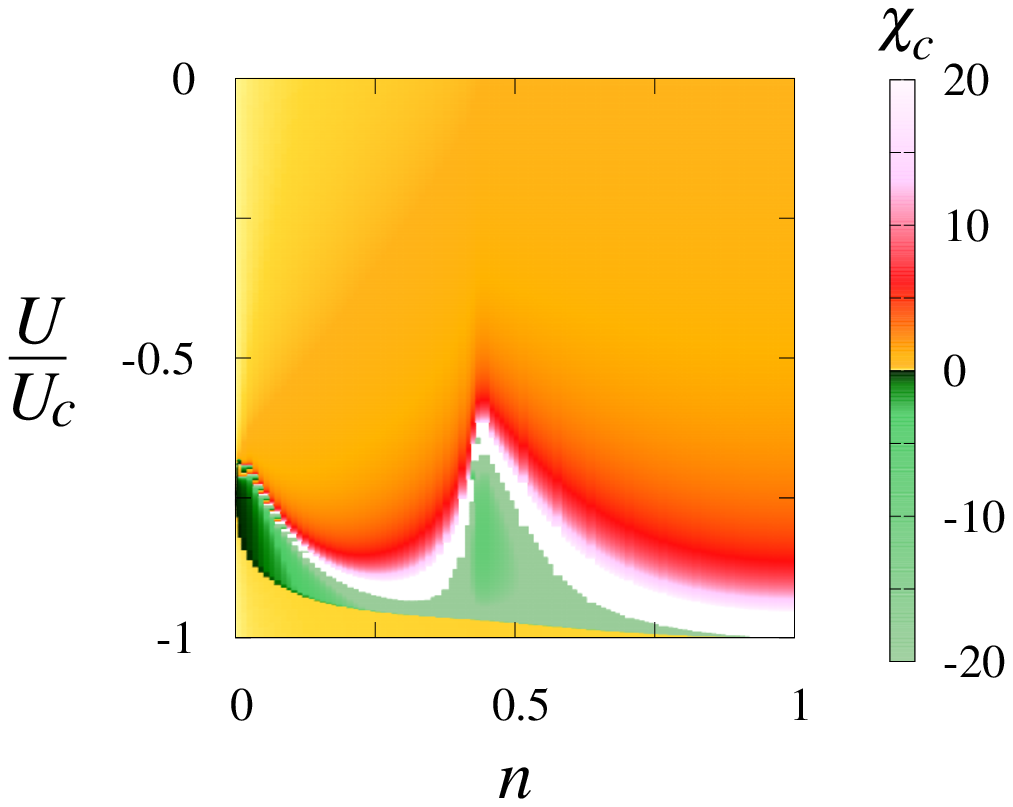}
\end{tabular}
    \caption{(left) Landau parameter $F_0^s$ and (right) static homogeneous
      charge susceptibility $\chi_c$ for $V= 0$, as functions of density $n$
      and onsite Coulomb interaction $U$. The white shade corresponds to
      values larger than 20.} 
    \label{fig:instab_0} 
\end{figure}

The stability of the paramagnetic state obtained in the previous section is 
investigated by means of the symmetric Landau Fermi-liquid parameter $F_0^s$
and the static charge response function $\chi_c$ in the long wavelength limit.
The latter quantities are numerically calculated within the one-loop approximation. 
For the present discussion, we consider an isotropic intersite Coulomb 
interaction restricted to the nearest neighbor sites, with $V_{ij}=V$.

\begin{figure}
   \centering
\begin{tabular}{cc}
\includegraphics[clip=true,trim={1.2cm 0 0cm 0}, width=0.55\textwidth]{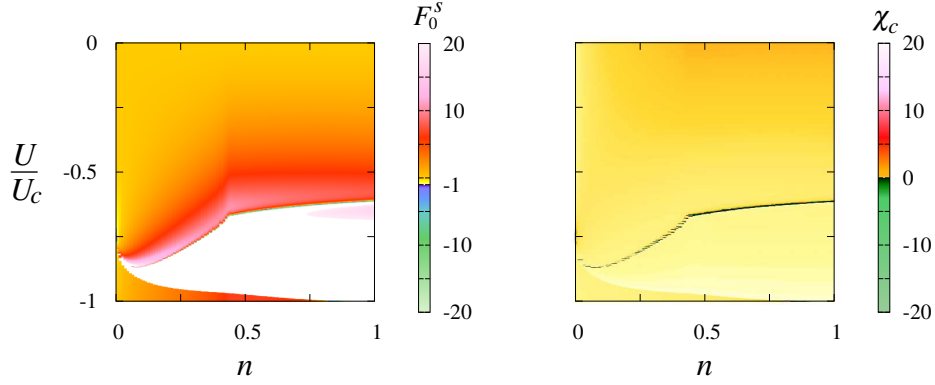}&
\includegraphics[clip=true,trim={3.3cm 0 0 0}, width=0.45\textwidth]{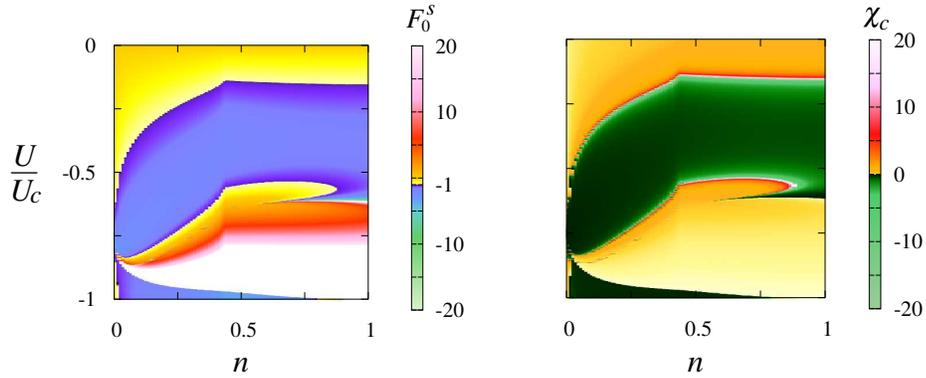}
\end{tabular}
    \caption{(left) Landau parameter $F_0^s$ and (right) static homogeneous
      charge susceptibility $\chi_c$ for $V= -0.2\ U$, as functions of density
      $n$ and onsite Coulomb interaction $U$. The white shade corresponds to
      values larger than 20.}
    \label{fig:instab_MP2} 
\end{figure}

\begin{figure}
   \centering
\begin{tabular}{cc}
\includegraphics[clip=true,trim={1cm 0 0cm 0}, width=0.55\textwidth]{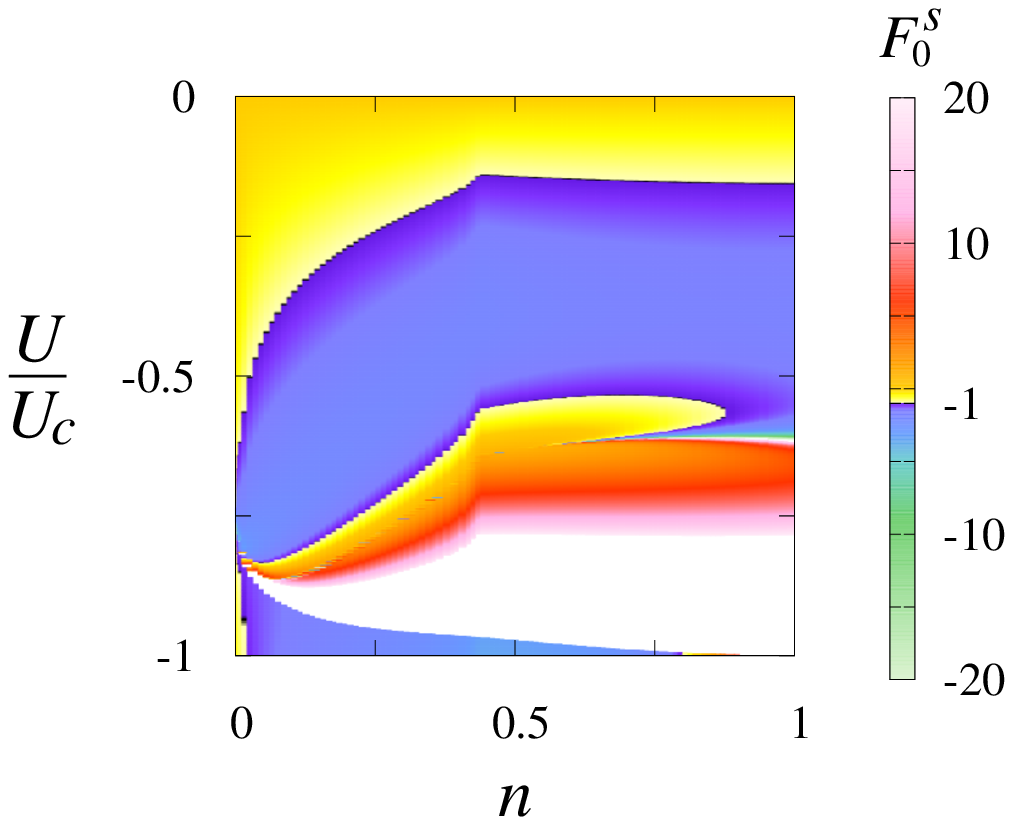}&
\includegraphics[clip=true,trim={3.3cm 0 0 0}, width=0.45\textwidth]{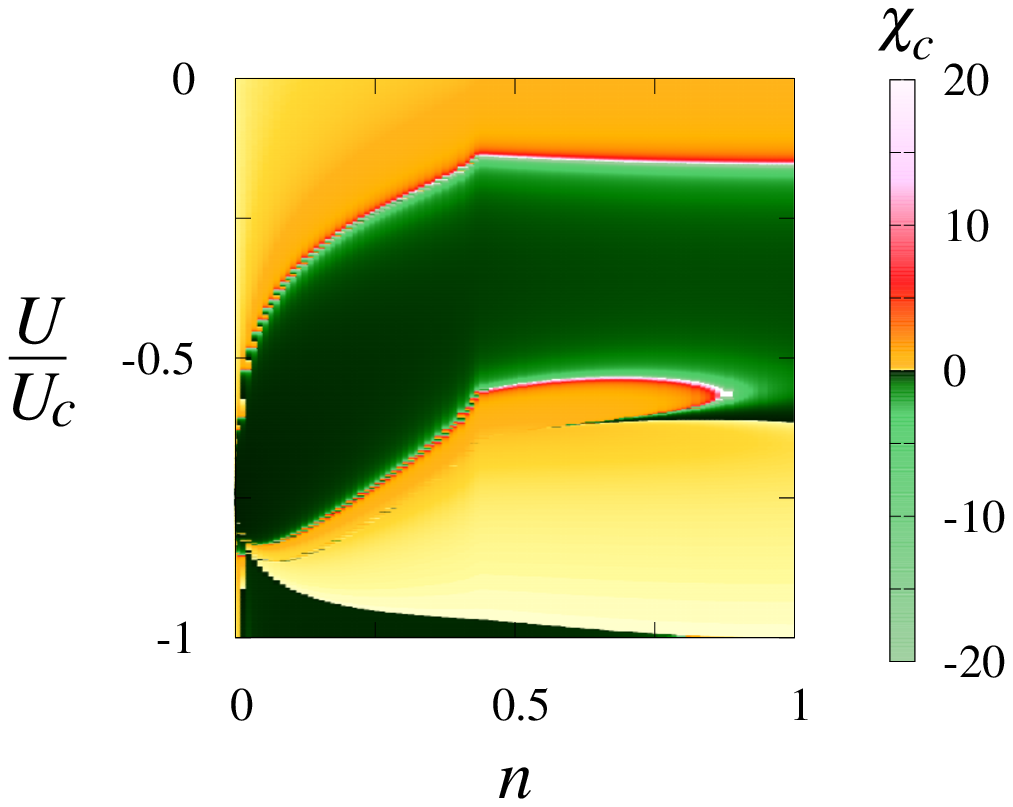}
\end{tabular}
    \caption{(left) Landau parameter $F_0^s$ and (right) static homogeneous
      charge susceptibility $\chi_c$ for $V= 0.2\ U$, as functions of density
      $n$ and onsite Coulomb interaction $U$. The white shade corresponds to
      values larger than 20.}
    \label{fig:instab_P2} 
\end{figure}

Figs.~\ref{fig:instab_0}--\ref{fig:instab_P2} display the variations of $F_0^s$ and 
$\chi_c$, as functions of the density $n$ and the onsite interaction $U$, for 
respectively $V=0$, $V = -0.2\ U$, and $V = 0.2\ U$. 
The shades in the graphics have been chosen to highlight the loci in the parameter 
space where $F_0^s = -1$. The values slightly larger than -1 are plotted with a light 
yellow hue, in contrast to the values slightly lower than -1 plotted with a dark blue 
hue. The value $F_0^s <-1$ results in a charge response $\chi_c <0$ (plotted with a 
green shade), which indicates an instability of the paramagnetic solution. 

For an attractive onsite coupling, without intersite interaction 
(see Fig.~\ref{fig:instab_0}), the uniform paramagnetic state is unstable in a narrow 
two-dimensional parameter space centered around quarter filling and $U \simeq 0.8\ U_c$.
Note that the instability arises within the region of existence of the saddle-point 
solution $x_+$, at values of $|U|<\tilde{U_c}$ well inside the Fermi liquid phase.

When a repulsive intersite interaction is turned on, the narrow window of
instability rapidly shrinks down to the vicinity of a critical point located
at $n=0$ and $U=-0.5\ U_0$. However, as shown in Fig.~\ref{fig:instab_MP2}, a
second region of instability emerges in the form of a line where $F_0^s$
diverges. The line of instability is located inside the Fermi
liquid phase. 

When both the onsite and intersite interactions are attractive, the
two-dimensional region of instability grows and is displaced towards lower
values of $|U|$ (see Fig.~\ref{fig:instab_P2}). The line of instability is
also present. We have noted that the location of this line seems not to depend
on the sign of the intersite interaction.

\section{Discussion and summary}

Summarizing, we have investigated the paramagnetic phase of the extended
attractive Hubbard model on the cubic lattice within the spin rotation invariant
Kotliar-Ruckenstein slave-boson representation, first on the saddle-point
level, and second at one-loop order. The  saddle-point equations may be cast
into a single algebraic equation, \cite{Vol87,FW} independent of the
non-local interaction parameters \cite{Lhou15}. In the repulsive case, it
reproduces the 
Brinkman-Rice metal-to-insulator transition at half-filling. Despite the
multiple solutions of the saddle-point equations no phase transition is found
away from half-filling, regardless of whether the density or the interaction
strength is varied (at zero temperature). In the attractive case we found the
situation to be qualitatively different, with a first order transition for any
density when increasing $-U$ from a Fermi liquid phase to a marginally
relevant one. In fact, the phase boundary is given by the line where the
effective mass diverges or, in other words, where the quasiparticle residue
vanishes. 

We then turned to the Landau parameter $F_0^s$ and, in particular, to its
dependence on the non-local interaction parameters. Again, we obtain results
that qualitatively differ from the repulsive case. Of course, the nature of
the instabilities differs, but one may wish to unify their
discussion. Intuitively, while one here expects some form of charge (pair)
density wave, the repulsive model naturally yields magnetic
instabilities/orderings. In particular, for a given density, the Landau
parameter $F_0^a$ may reach $-1$ at a critical value of the interaction
strength, and remains smaller than $-1$ for all values of the interaction
strength larger than the critical one. This is indeed yielded by the calculation
\cite{Doll1}. That does not prove that the ground state is ferromagnetic, as
these instabilities compete with incommensurate ones. And, indeed, a
ferromagnetic ground state could at best be stabilized for very large $U$ and
small doping \cite{Doll2}, in agreement with Nagaoka
ferromagnetism. Nevertheless, for a given point in the phase diagram, $F_0^a <
-1$ remains a rather reliable indicator of a magnetic instability. In contrast,
for the attractive cases studied here, the obtained behavior of $F_0^s$ differs
qualitatively. For instance, for vanishing non-local interaction parameters,
we found $F_0^s < -1$ in a rather narrow window centered around quarter
filling and $U \simeq 0.8\ U_c$. This window is even far narrower for 
$V = -0.2\ U$ and is shifted towards smaller values of $|U|$. In fact, the
expected pair 
density wave is not signaled by homogeneous instabilities. Hence, contrary
to the repulsive model,  $F_0^s < -1$ does not turn out to be a good indicator
of incommensurate pair density wave instabilities, which determination remains
as a task for future work.
 
\section*{Acknowledgments}

R.F. is grateful for the warm hospitality at MPI Stuttgart where part of this
work has been done.

\end{document}